\documentclass[aps,superscriptaddress,twocolumn,showpacs]{revtex4-1}
\pdfoutput=1
\usepackage{color}
\usepackage{amsmath, amsthm, amsfonts}    
\usepackage{amssymb}
\usepackage{mathtools}
\usepackage{mathptmx} 
\usepackage[table]{xcolor}
\usepackage{dsfont}
\usepackage{enumitem} 
\usepackage{appendix}
\usepackage{braket}
\usepackage{pifont}   

\usepackage[skins,theorems]{tcolorbox}
\tcbset{highlight math style={enhanced,
  colframe=red,colback=white,arc=0pt,boxrule=1pt}}
\usepackage{cancel}
\usepackage{empheq}
\usepackage{lipsum}
\usepackage{graphicx}
\usepackage{tikz}
\usepackage{makecell}
\usepackage[unicode=true,pdfusetitle, bookmarks=true,bookmarksnumbered=false,bookmarksopen=false,
 breaklinks=true,pdfborder={0 0 0},backref=false,colorlinks=true,citecolor=blue]{hyperref}
\usepackage{graphicx}   
\usepackage[caption=false]{subfig}       
\usepackage{bm}            
\usepackage[normalem]{ulem}  

\newlength\imagewidth
\newlength\imagescale
\def\be{\begin{eqnarray}}
\def\ee{\end{eqnarray}}

\newcommand{\til}[1]{{\overline{#1}}}

\def\p{{\bf p}}
\def\m{{\bf m}}

\usepackage{calrsfs}
\DeclareMathAlphabet{\mathcal}{OMS}{cmsy}{m}{n}
\def\im{{\rm i}}

\definecolor{JOT-color}{named}{blue}
\definecolor{CSF-color}{named}{orange}
\newtcolorbox{keyresultbox}{
  colback=white!10, 
  colframe=black,  
  boxrule=0.5pt,
  arc=2pt,         
  left=6pt, right=6pt, top=6pt, bottom=6pt
}

\begin{document}

\title{A Universal Magnetoelectric Limit for Chiral and Tellegen Bi-Isotropic Scatterers}

\author{Jorge Olmos-Trigo}
\email{jolmostrigo@gmail.com}
\affiliation{Departamento de Física de Materiales $\&$ Condensed Matter Physics Center ($ \rm IFIMAC$), Universidad Autónoma de Madrid, 28049
Madrid, Spain.}

\begin{abstract}
We reveal the existence of a universal upper bound on the magnetoelectric coupling of any bi-isotropic nanoparticle. The bound arises solely from energy conservation, making it independent of the specific material properties of the nanoparticle and illumination conditions. Moreover, the bound does not rely on reciprocity, being identical for reciprocal (chiral) and non-reciprocal (Tellegen) nanoparticles.
We further show that the chiral Mie coefficient of spherical particles of arbitrary optical size obeys the same bound across all multipolar scattering channels. Our results introduce a universal metric on the magnetoelectric coupling of bi-isotropic objects, setting identical limits on chiral and Tellegen light-matter interactions at the single particle level.
\end{abstract}

\maketitle
{\emph{Introduction.—}}  
Bi-isotropic nanoparticles (NPs) exhibit magnetoelectric coupling, whereby the electric and magnetic responses of a material system can be excited by both the electric and magnetic components of the electromagnetic field. This magnetoelectric coupling is often classified, according to reciprocity, into two categories: reciprocal (chiral)~\cite{biot1845instructions,pasteur1848memoires} and non-reciprocal, as in the case of the Tellegen gyrator~\cite{tellegen1948gyrator}.

Most studies have focused on chiral NPs, whose response depends on the handedness of light, often referred to as electromagnetic helicity~\cite{calkin1965invariance, fernandez2013electromagnetic, olmos2020unveiling, ozaktas2025all, olmos2024revealing, olmos2024solving}.  Chiral light--matter interactions underlie a broad range of nanophotonic phenomena involving electromagnetic helicity, including enantioselective  forces~\cite{zhao2017nanoscopic, jin2024harnessing, genet2022chiral, rukhlenko2016completely, wang2014lateral, hayat2015lateral, zhao2016enantioselective, zhang2017all, man2024construction, zhang2024enantioselective, kamandi2017enantiospecific, shi2020optical, li2020enantioselective, martinez2024chiral, chen2014tailoring, sifat2022force, xiong2024enantioselective, serrera2026enhanced, pellegrini2019superchiral, champi2019optical, canaguier2016plasmonic, canaguier2015chiral, hou2021separating, fang2021optical,yao2024sorting, martinez2025chiral, zhang2025high}, chiral torques~\cite{liu2019separation, chen2017optical, rahimzadegan2016optical, wen2025optical, zhang2019optical, vovk2017chiral}, chiral scattering~\cite{shang2013analysis}, and circular dichroism enhancements~\cite{paul2025enhanced, vestler2019enhancement, mohammadi2023nanophotonic, tang2010optical, mohammadi2024nanophotonic, sadrara2023large, kim2022enantioselective, hanifeh2020optimally}.

Non-reciprocal magnetoelectric effects have also attracted growing attention~\cite{bliokh2014magnetoelectric, saltykova2025analytical, safaei2024optical}. Particularly, the hallmark of the Tellegen response~\cite{tellegen1948gyrator}, a Kerr rotation in reflection while the transmitted polarization remains unchanged, has been detected~\cite{yang2025gigantic, jazi2025realization}, opening routes for axion electrodynamics.

\begin{figure}
    \centering
\includegraphics[width=1\linewidth]{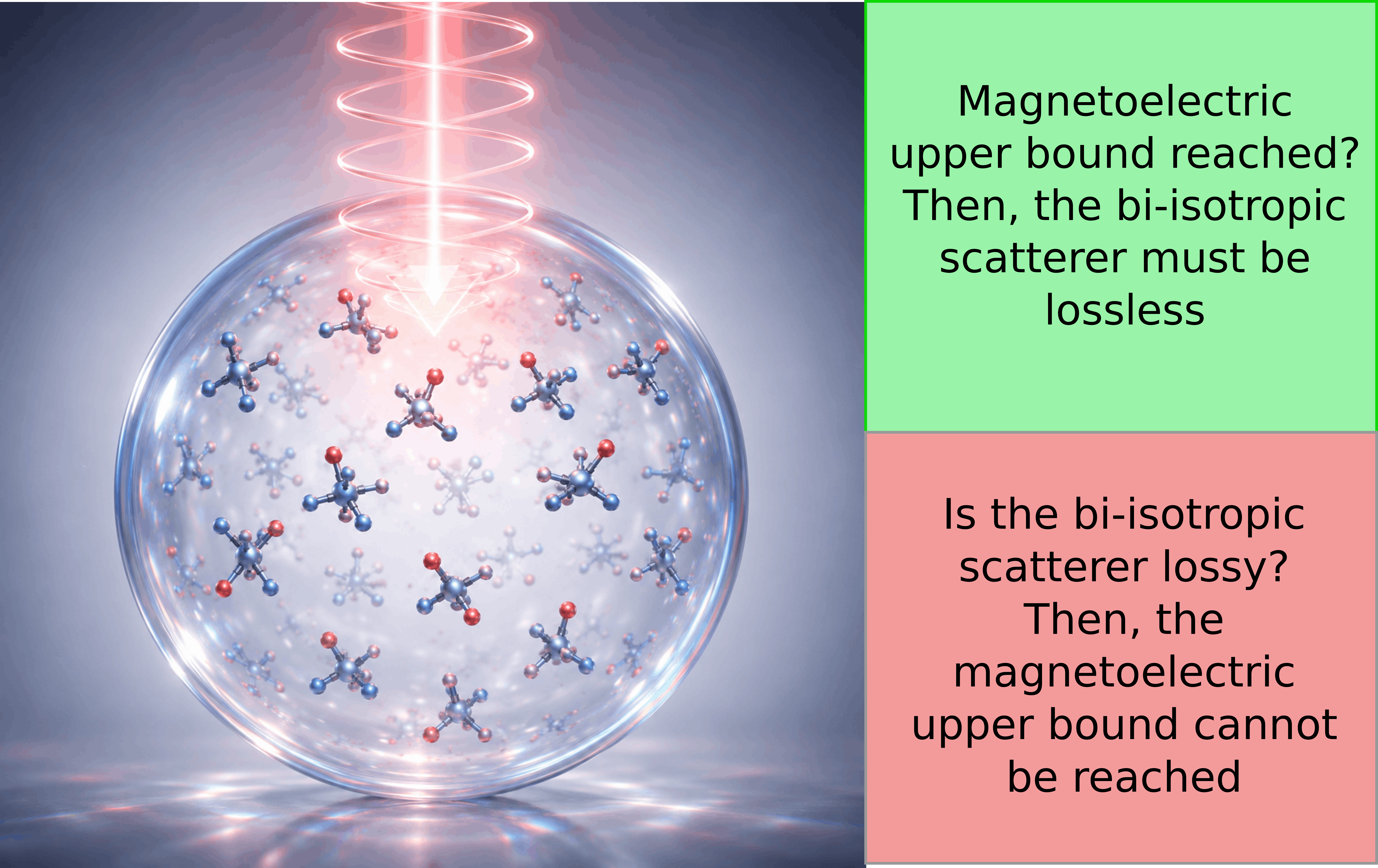}
    \caption{Sketch and the implications of the magnetoelectric limit.}
\label{fig:placeholder}
\end{figure}

Despite the rich diversity of the light-scattering systems studied in Refs.~\cite{zhao2017nanoscopic, jin2024harnessing, genet2022chiral, rukhlenko2016completely, wang2014lateral, hayat2015lateral, zhao2016enantioselective, zhang2017all, man2024construction, zhang2024enantioselective, kamandi2017enantiospecific, shi2020optical, li2020enantioselective, martinez2024chiral, chen2014tailoring, sifat2022force, xiong2024enantioselective, serrera2026enhanced, pellegrini2019superchiral, champi2019optical, canaguier2016plasmonic, canaguier2015chiral, hou2021separating, fang2021optical,yao2024sorting, martinez2025chiral, zhang2025high,   liu2019separation, chen2017optical, rahimzadegan2016optical, wen2025optical, zhang2019optical, vovk2017chiral, shang2013analysis,paul2025enhanced,vestler2019enhancement, mohammadi2023nanophotonic, tang2010optical, mohammadi2024nanophotonic, sadrara2023large, kim2022enantioselective, hanifeh2020optimally, bliokh2014magnetoelectric,  saltykova2025analytical, safaei2024optical, yang2025gigantic, jazi2025realization}, the underlying physics of all these phenomena relies on two complex-valued scalars: the cross-polarizabilities $\alpha_{\rm em}$ and $\alpha_{\rm me}$. In particular, the strength of chiral and Tellegen light-matter interactions is determined by these quantities. For instance, when $\alpha_{\rm em}=\alpha_{\rm me}=0$, the NP reduces to a purely electric-magnetic dipolar scatterer, whose upper bounds are well established~\cite{albaladejo2010radiative}. In contrast, no general upper bounds are known for $\alpha_{\rm em}$ and $\alpha_{\rm me}$, and it remains unclear whether such limits exist at all. Establishing them would determine the ultimate achievable magnetoelectric coupling in any bi-isotropic NP, thereby defining the fundamental limits of chiral and Tellegen light-matter interactions.
In this work, we reveal a universal, identical upper bound for $\alpha_{\rm em}$ and $\alpha_{\rm me}$. We derive a simple expression for this limit and find that $|\alpha_{\rm em}|, |\alpha_{\rm me}| \le 3\pi/k^3$, i.e., half of the maximum value of the polarizability of a resonant isotropic point electric dipole. In addition, we demonstrate that when the magnetoelectric bound is reached for NPs meeting $\alpha_{\rm em} = \pm \alpha_{\rm me}$, corresponding to chiral ($-$) or Tellegen ($+$) responses, such NPs must be lossless. In other words, lossy bi-isotropic NPs cannot attain the upper bound. All this information is summarized in Fig.~\ref{fig:placeholder}.

We then generalize our results to chiral spherical objects of arbitrary optical size using the T-matrix formalism~\cite{mishchenko2002scattering}. Specifically, we show that the chiral coefficients introduced by Prof.~C.~F.~Bohren in 1974~\cite{bohren1974light} are upper bounded and can reach at most one half of the maximum value of the electric scattering coefficient. 




{\emph{The Polarizability tensor of any Bi-Isotropic NP.—--}}
The electromagnetic response of a small NP can be described by its induced electric dipolar moment $\p$ and magnetic dipolar moment $\m$~\cite{nieto2011angle}. In the case of bi-anisotropic NPs, each of these dipolar moments can be excited by both the incident electric field $\mathbf{E}$ and the magnetic field $\mathbf{H}$. This coupling is featured by the $6 \times 6$ polarizability tensor $\tilde{\alpha}$~\cite{fernandez2016objects}:
\begin{equation} \label{Rels_gen}
\begin{pmatrix}
\mathbf{p} \\
\mathbf{m}
\end{pmatrix}
=
\tilde{\alpha}
\begin{pmatrix}
\mathbf{E} \\
\mathbf{H}
\end{pmatrix}
=
\begin{pmatrix}
\til{\alpha}_{\rm ee} & \til{\alpha}_{\rm em} \\
\til{\alpha}_{\rm me} & \til{\alpha}_{\rm mm}
\end{pmatrix}
\begin{pmatrix}
\mathbf{E} \\
\mathbf{H}
\end{pmatrix}.
\end{equation}
Here, $\til{\alpha}_{\mathrm{ee}}$ and $\til{\alpha}_{\mathrm{mm}}$ denote the electric and magnetic polarizability tensors, respectively, and $\til{\alpha}_{\mathrm{em}}$ and $\til{\alpha}_{\mathrm{me}}$ denote the electro-magnetic and magneto-electric polarizability tensors, respectively. 
For bi-anisotropic NPs~\cite{alaee2015all}, each $\til{\alpha}_{\rm ij}$ is a $3\times3$ complex-valued tensor. However, in the case of bi-isotropic NPs, rotational symmetry reduces each of these polarizability tensors to scalars. That is~\cite{olmos2025novel}, 
\begin{align} \nonumber
\til{\alpha}_{\rm ee} = \alpha_{\rm ee} \rm{I}_{3},  && \til{\alpha}_{\rm mm} = \alpha_{\rm mm} \rm{I}_{3}, && \til{\alpha}_{\rm em} = \alpha_{\rm em} \rm{I}_{3}, && 
\til{\alpha}_{\rm me} = \alpha_{\rm me} \rm{I}_{3},
\end{align}
where
$\rm{I}_3$ is the $3 \times 3$ identity matrix~\footnote{In our framework, all polarizabilities are complex-valued scalars with dimensions of volume. Moreover, a harmonic time dependence $e^{-\mathrm{i}\omega t}$ is assumed and omitted from the notation.  In this regard, the ${\alpha}_{\rm{ij}}$ polarizabilities are frequency dependent, but we only write such dependence later.}.

In this bi-isotropic setting, Eq.~\eqref{Rels_gen} simplifies to:
\begin{equation} \label{Pols}
\begin{pmatrix}
\mathbf{p} \\
\mathbf{m}
\end{pmatrix}
=
\begin{pmatrix}
\alpha_{\rm ee} & \alpha_{\rm em} \\
\alpha_{\rm me} & \alpha_{\rm mm}
\end{pmatrix}
\begin{pmatrix}
\mathbf{E} \\
\mathbf{H}
\end{pmatrix}.
\end{equation}
Equation~\eqref{Pols} is ubiquitous in modeling the interaction between the electromagnetic field and chiral and Tellegen light-matter interactions. In fact, it constitutes the standard framework employed in Refs~\cite{zhao2017nanoscopic, jin2024harnessing, genet2022chiral, rukhlenko2016completely, wang2014lateral, hayat2015lateral, zhao2016enantioselective, zhang2017all, man2024construction, zhang2024enantioselective, kamandi2017enantiospecific, shi2020optical, li2020enantioselective, martinez2024chiral, chen2014tailoring, sifat2022force, xiong2024enantioselective, serrera2026enhanced, pellegrini2019superchiral, champi2019optical, canaguier2016plasmonic, canaguier2015chiral, hou2021separating, fang2021optical,yao2024sorting, martinez2025chiral, zhang2025high,   liu2019separation, chen2017optical, rahimzadegan2016optical, wen2025optical, zhang2019optical, vovk2017chiral, shang2013analysis,paul2025enhanced,vestler2019enhancement, mohammadi2023nanophotonic, tang2010optical, mohammadi2024nanophotonic, sadrara2023large, kim2022enantioselective, hanifeh2020optimally, bliokh2014magnetoelectric,  saltykova2025analytical, safaei2024optical, yang2025gigantic, jazi2025realization} and the one we use in the first part of this work. In particular, we derive next the exact energy conservation law governing the polarizabilities $\alpha_{\rm ee}$, $\alpha_{\rm mm}$, $\alpha_{\rm em}$, and $\alpha_{\rm me}$. As shown below, this law is essential for establishing the upper bounds on the cross-polarizabilities $\alpha_{\rm em}$ and $\alpha_{\rm me}$.

\emph{A Universal Bound for Any Bi-Isotropic NP.--—}  For any dipolar NP, bi-anisotropic or not, the energy conservation law can be written in terms of $\tilde{\alpha}$ as~\cite[Eq.~(10)]{belov2003condition}:
\begin{equation}\label{belov}
\frac{\tilde{\alpha}-\tilde{\alpha}^\dagger}{2\mathrm{i}}
\ge \frac{k^3}{6\pi}\tilde{\alpha}^\dagger \tilde{\alpha},
\end{equation}
where $``\dagger"$ denotes Hermitian conjugation.
The left-hand side of Eq.~\eqref{belov} is proportional to extinction, whereas the right-hand side is proportional to scattering.
For any lossless NP, extinction equals scattering, yielding zero absorption.

 At this point, we express
Eq.~\eqref {belov} in terms of  
$\alpha_{\rm ee}$, $\alpha_{\rm mm}$, $\alpha_{\rm em}$, and $\alpha_{\rm me}$. This  can be done by introducing the Hermitian positive semidefinite (HPSD)  absorbance matrix $\tilde{\mathcal{A}}$~\cite{albaladejo2010radiative, golat2026dipole, olmos2026exact}:
\begin{equation} \label{A_abs}
\tilde{\mathcal{A}}=
\frac{\tilde{\alpha}-\tilde{\alpha}^\dagger}{2{\rm i}}
-\frac{k^3}{6\pi}\tilde{\alpha}^\dagger\tilde{\alpha}
= \begin{pmatrix} \mathcal{A}_{11} & \mathcal{A}_{12} \\ \mathcal{A}_{12}^* & \mathcal{A}_{22} \end{pmatrix}.
\end{equation}
The matrix elements of $\tilde{\mathcal{A}}$ can be calculated using the $2 \times 2$ bi-isotropic polarizability tensor given in Eq.~\eqref{Pols}, yielding:
\begin{align} \label{A11}
\mathcal{A}_{11} &= \Im\{\alpha_{\rm ee}\} - \frac{k^3}{6\pi}\left(|\alpha_{\rm ee}|^2 + |\alpha_{\rm em}|^2\right),\\ \label{A22}
\mathcal{A}_{22} &= \Im\{\alpha_{\rm mm}\} - \frac{k^3}{6\pi}\left(|\alpha_{\rm mm}|^2 + |\alpha_{\rm me}|^2\right),\\ \label{A12}
\mathcal{A}_{12} &= \frac{\alpha_{\rm em} - \alpha_{\rm me}^*}{2{\rm i}} 
- \frac{k^3}{6\pi}\left(\alpha_{\rm ee}^*\alpha_{\rm em} + \alpha_{\rm me}^*\alpha_{\rm mm}\right).
\end{align}
At this point, we use the fact that $\tilde{\mathcal{A}}$ is, by construction a HPSD matrix. In particular, we use  Sylvester’s criterion~\cite{gilbert1991positive}, which, as a reminder to the reader, imposes that the principal minors of any HSPD matrix must be nonnegative. Applying Sylvester’s criterion to $\tilde{\mathcal{A}}$ given in Eq.~\eqref{A_abs} yields
\begin{equation} \label{sylver}
\mathcal{A}_{11}\ge 0, \qquad
\mathcal{A}_{22}\ge 0, \qquad
\mathcal{A}_{11}\mathcal{A}_{22} \ge |\mathcal{A}_{12}|^2 .
\end{equation}
It is important to note that any physically admissible polarizability model of a bi-isotropic NP must satisfy Eq.~\eqref{sylver}.
As a sanity check, we note that for $\alpha_{\rm{em}} = \alpha_{\rm{me}} = 0$, one recovers the well-known relations for isotropic NPs~\cite{olmos2019enhanced, olmos2019asymmetry} in the dipolar regime, namely, 
\begin{align} \nonumber
\Im \{ \alpha_{\rm{ee}} \} \ge \frac{k^3}{6 \pi}  |\alpha_{\rm{ee}}|^2, &&  \Im \{ \alpha_{\rm{mm}} \} \ge \frac{k^3}{6 \pi} |\alpha_{\rm{mm}}|^2. 
\end{align}
Moreover, if the NP is lossless, then we have $\mathcal{A}_{11} = \mathcal{A}_{22} = \mathcal{A}_{12} = 0$ (and the converse statement also holds).
In this case, the following relations are satisfied:
\begin{align} \nonumber
\Im\{\alpha_{\rm ee}\} &= \frac{k^3}{6\pi}\left(|\alpha_{\rm ee}|^2 + |\alpha_{\rm em}|^2\right),\\ \nonumber
\Im\{\alpha_{\rm mm}\} &= \frac{k^3}{6\pi}\left(|\alpha_{\rm mm}|^2 + |\alpha_{\rm me}|^2\right),\\ \nonumber
 \frac{\alpha_{\rm em} - \alpha_{\rm me}^*}{2{\rm i}} 
&= \frac{k^3}{6\pi}\left(\alpha_{\rm ee}^*\alpha_{\rm em} + \alpha_{\rm me}^*\alpha_{\rm mm}\right).
\end{align}
We stress that any bi-isotropic NP must fulfill the previous identities in the absence of losses.

Next, we turn our attention to the first two Sylvester minors given by Eq.~\eqref{sylver}, which can be  rewritten as:
\begin{align} \label{upper_e}
\frac{k^3}{6\pi} |\alpha_{\rm em}|^2 &\le 
\Im\{\alpha_{\rm ee}\} - \frac{k^3}{6\pi}|\alpha_{\rm ee}|^2,\\
\label{upper_m}
\frac{k^3}{6\pi} |\alpha_{\rm me}|^2 &\le 
\Im\{\alpha_{\rm mm}\} - \frac{k^3}{6\pi}|\alpha_{\rm mm}|^2.
\end{align}
Equations~\eqref{upper_e}-\eqref{upper_m} impose the following constrains: if $\alpha_{\rm{ee}} = 0$ ($\alpha_{\rm{mm}} = 0$), then $\alpha_{\rm{em}} = 0$ $(\alpha_{\rm{me}} = 0$), in close agreement with previous findings~\cite{albooyeh2016purely}. Having noted these well-known limits, we now maximize the
right-hand side of Eqs.~\eqref{upper_e}-\eqref{upper_m} to determine the overlooked upper bounds of $ |\alpha_{\rm{em}}|^2$ and $ |\alpha_{\rm{me}}|^2$.

Let $z$ denote either $\alpha_{\rm ee}$ or $\alpha_{\rm mm}$, and write
$z'=\Re\{z\}$ and $z''=\Im\{z\}$. The relevant function to be maximized in both electric and magnetic dipolar channels is then:
\begin{equation} \nonumber
f(z)=\Im\{z\}-\frac{k^3}{6\pi}|z|^2
= z''-\frac{k^3}{6\pi}\left(z'^2+z''^2\right).
\end{equation}
The function $f(z)$ attains its global maximum, given by $3 \pi /(2k^3)$, at
$z'=0$ and $z''=3\pi/k^3$. We next show that this complex-valued point ($z'=0, z''=3\pi/k^3$) lies within the  admissible domain of both
$\alpha_{\rm ee}$ and $\alpha_{\rm mm}$ and is physically realizable. To show this, we write $\mathcal{A}_{11} \ge 0$ and $\mathcal{A}_{22} \ge 0$ as: 
\begin{align}
\Im\{\alpha_{\rm ee}\} &\ge \frac{k^3}{6\pi}\left(|\alpha_{\rm ee}|^2 + |\alpha_{\rm em}|^2\right) \ge   \frac{k^3}{6\pi}|\alpha_{\rm ee}|^2,  \\
\Im\{\alpha_{\rm mm}\} &\ge \frac{k^3}{6\pi}\left(|\alpha_{\rm mm}|^2 + |\alpha_{\rm me}|^2\right) \ge   \frac{k^3}{6\pi}|\alpha_{\rm mm}|^2.  
\end{align}
From these inequalities, it follows that the allowed values of 
$\alpha_{\rm ee}$ and $\alpha_{\rm mm}$ are unmodified by the presence of $\alpha_{\rm em}$ and $\alpha_{\rm me}$ and are given by $\Im\{z\} \ge (k^3/ 6 \pi) |z|^2$ with $\Im\{z\} \ge 0$ for $z=\alpha_{\rm ee},\alpha_{\rm mm}$. This condition defines a closed region in the complex plane, which can be written as
\begin{equation} \label{closed}
z'^2 + \left(z'' - \frac{3\pi}{k^3}\right)^2 \le \left(\frac{3\pi}{k^3}\right)^2.
\end{equation}
It can be noted that the complex-valued point maximizing $f(z)$, given by $z'=0$ and $z''=3\pi/k^3$, lies inside the domain given by Eq.~\eqref{closed}. Therefore, it is physically admissible and the corresponding maximum 
$f_{\max} = 3\pi/(2k^3)$ is attainable. Taking into account all this information, we can finally write
\begin{align} \label{ultimate_em_bound}
|\alpha_{\rm em}| &\leq \frac{3\pi}{k^3}, 
&\quad |\alpha_{\rm em}| = \frac{3\pi}{k^3} &\Longrightarrow \alpha_{\rm ee} = \im \frac{3\pi}{k^3}, \\ \label{ultimate_me_bound}
|\alpha_{\rm me}| &\leq \frac{3\pi}{k^3}, 
&\quad |\alpha_{\rm me}| = \frac{3\pi}{k^3} &\Longrightarrow \alpha_{\rm mm} = \im \frac{3\pi}{k^3}.
\end{align}
Equations~\eqref{ultimate_em_bound}–\eqref{ultimate_me_bound} represent the first key result of this work; $\alpha_{\rm em}$ and $\alpha_{\rm me}$ share the same ultimate upper bound. This upper bound is universal: regardless of the reciprocal or material properties of the NPs, the incident illumination, or the observation point (near- or far-field), both $|\alpha_{\rm em}|$ and $|\alpha_{\rm me}|$ are limited to $3\pi/k^3$, exactly half of the maximum value of the polarizability of a resonant isotropic point-electric dipole. 

Importantly, the saturation of these upper bounds is not an independent condition on the cross-couplings alone. Instead, it reflects a fundamental constraint of the full dipolar response: when $|\alpha_{\rm em}|$ or $|\alpha_{\rm me}|$ reaches $3\pi/k^3$, the direct electric (magnetic) polarizabilities are purely imaginary with half of their absolute maximum value.

Next, we examine the features of chiral and Tellegen NPs when the right-hand side of Eqs.~\eqref{ultimate_em_bound}-\eqref{ultimate_me_bound} is met.


\emph{Features of chiral and Tellegen NPs at the upper bound.}
The first key observation is that both chiral and Tellegen NPs, characterized by $\alpha_{\rm em} = -\alpha_{\rm me}$ and $\alpha_{\rm em} = +\alpha_{\rm me}$, respectively, must be lossless at their upper bound. To demonstrate this feature, we notice that the right-hand side of Eq.~\eqref{ultimate_em_bound} implies $\mathcal{A}_{11}=0$, while the right-hand side of Eq.~\eqref{ultimate_me_bound} yields $\mathcal{A}_{22}=0$. Since the right-hand side of Eq.~\eqref{sylver} requires $\mathcal{A}_{11}\mathcal{A}_{22} \ge |\mathcal{A}_{12}|^2$, it follows necessarily that $\mathcal{A}_{12}=0$. We stress that $\mathcal{A}_{11} = \mathcal{A}_{22} = |\mathcal{A}_{12}| = 0$ implies a lossless NP (and the converse also holds). We can now encompass all this information as:
\be 
\label{Lossless}
|\alpha_{\rm em}| &=& |\alpha_{\rm me}| =  \frac{3\pi}{k^3}
\Longrightarrow
\text{Lossless  NPs.}
\ee

At this point, we contextualize our results with previous ones, delving into energy conservation in bi-anisotropic light-matter interactions. Previous studies have argued that in small passive NPs, the magnetoelectric coupling cannot exceed the direct polarization effects, leading to the inequality~\cite{sersic2011magnetoelectric}:
\begin{equation} \label{sersic}
|\alpha_{\rm{em}}\alpha_{\rm{me}}| \le |{\alpha_{\rm{ee}} \alpha_{\rm{mm}}}|.
\end{equation}
In the following, we show that the upper bound $|\alpha_{\rm{em}}|$ and $\alpha_{\rm{me}}|$ implied by Eq.~\eqref{sersic} might lead to nonphysical values when applied to bi-isotropic NPs. To show this, 
we note that the right-hand side of Eq.~\eqref{sersic} is upper bounded, $|{\alpha_{\rm{ee}} \alpha_{\rm{mm}}}| \leq 36 \pi^2 / k ^6$. Thus, we are left with $
|\alpha_{\rm{em}}\alpha_{\rm{me}}| \le {36 \pi^2}/{k^6}$. 
At this point, we impose the left-hand side of Eqs.~\eqref{ultimate_em_bound}-\eqref{ultimate_me_bound} yielding
$
|{\alpha_{\rm{em}} \alpha_{\rm{me}}}| \leq 9 \pi^2 / k ^6$. By comparing both limits, we conclude that the upper bounds of $\alpha_{\rm{em}}$ and $\alpha_{\rm{me}}$ implied by Eq.~\eqref{sersic} overestimate the actual allowed values for bi-isotropic NPs, as we previously anticipated. In particular, the values $9 \pi/k^6 \leq |{\alpha_{\rm{em}} \alpha_{\rm{me}}}| \leq 36 \pi/k^6$ are simple not allowed by energy conservation.

Hitherto, we have restricted our discussion to dipolar bi-isotropic NPs. We now show that the universal character of the magnetoelectric bounds extends far beyond the dipolar regime, applying also to chiral spherical objects of arbitrary optical size. It is important to mention that these systems have attracted considerable interest in recent years~\cite{ali2020enantioselective, ali2020probing, li2022enatioselective, wu2022selective, dutra2026circular, shi2020chirality, lai2024observation, feng2025revealing, miao2025long, shi2026self, diniz2025probing, olmos2024spheres, zheng2021selective}.

To analyze these chiral objects, we employ the transition-matrix (T-matrix) formalism, which fully characterizes the electromagnetic response of an arbitrary object under general illumination conditions~\cite{mishchenko2002scattering,asadova2025t}.

\emph{The T-matrix of any chiral spherical object.—--}
The T-matrix of a chiral spherical object can be expressed in terms of the Mie coefficients $a_\ell$, $b_\ell$, and $c_\ell$ as~\cite[Eq.~(12)]{patti2019chiral}:
\begin{equation} \label{patti}
T = \bigoplus_\ell T_\ell, 
\quad 
T_\ell = R_\ell \otimes I_{2\ell+1}, 
\quad
R_\ell= 
\begin{pmatrix} 
-a_\ell & \im c_\ell \\ 
\im c_\ell & -b_\ell 
\end{pmatrix}.
\end{equation}
Here $\bigoplus_\ell$ denotes the direct sum and $\otimes$ the tensor product.
Each block $T_\ell$ corresponds to a multipolar order $\ell$ and acts on the
$2(2\ell+1)$ angular momentum channels $m=-\ell,\dots,\ell$. 

For completeness, we also provide the explicit form of the Mie coefficients $a_\ell$, $b_\ell$, and $c_\ell$ as in [page~188]\cite{bohren2008absorption}:
\be \nonumber
a_\ell  &=& \Delta^{-1}_\ell  \left[  V_\ell (R)A_\ell (L) + V_\ell (L)A_\ell (R) \right], \\ \nonumber
b_\ell  &=& \Delta^{-1}_\ell  \left[ W_\ell (L)B_\ell (R) + W_\ell (R)B_\ell (L) \right], \\ \nonumber
c_\ell  &=& i \Delta^{-1}_\ell  \left[  W_\ell (R) A_\ell (L) - W_\ell (L)A_\ell (R) \right],
\ee
where $\Delta_\ell  =  W_\ell (L) V_\ell (R) + V_\ell (L)W_\ell (R)$. 
Here, we have defined the following auxiliary functions
\begin{subequations}
\be \nonumber
W_\ell (J) &=& \rm{m} \psi_\ell (m_J x) \xi^{'}_\ell (x)- \xi_\ell (x) \psi^{'}_\ell (m_J x), \\\nonumber
V_\ell (J) &=& \psi_\ell (\rm{m}_J x) \xi^{'}_\ell (x)- m\xi_\ell (x) \psi^{'}_\ell (m_J x), \\\nonumber
A_\ell (J) &=& \rm{m} \psi_\ell (m_J x) \psi^{'}_\ell (x)- \psi_\ell (x) \psi^{'}_\ell (m_J x), \\\nonumber
B_\ell (J) &=& \psi_\ell (\rm{m}_J x) \psi^{'}_\ell (x)- m\psi_\ell (x) \psi^{'}_\ell (m_J x). \label{ultimo}
\ee
\end{subequations}
Here $J = L, R$, with $\rm{m}_L  = \rm{m} - \chi$ and  $\rm{m}_R = \rm{m} + \chi$, $\rm{m}$ being the refractive index contrast,  $\chi$ is a parameter modeling the  intrinsic  chirality of the spherical object, ${\rm{x}} = ka = 2 
\pi (a/ \lambda) $ is the dimensionless size parameter of the particle,  $a$ and $\lambda$ being its radius and radiation wavelength, respectively. Moreover, $\psi_\ell  (z) = z j_\ell (z)$ and $\xi_\ell  (z) = z h^{(1)}_\ell (z)$ denote Riccati-Bessel functions~\cite{jackson1999electrodynamics}, $j_\ell$ being the spherical Bessel function and $h^{(1)}_\ell$ the spherical Hankel function of first kind~\cite{jackson1999electrodynamics}. 

In order to derive upper bound of $c_\ell$, we need to consider the full HPSD absorbance matrix, hereafter denoted by $\mathcal{Q}$. This $\mathcal{Q}$-matrix follows directly from the T-matrix and can be written as~\cite[Eq.~(10)]{le2013radiative}:
\[
\mathcal Q = -T - T^\dagger - 2 T^\dagger T.
\]
Expanding $\mathcal Q$ using Eq.~\eqref{patti}, we arrive at:
\[ \nonumber
\mathcal{Q} = \bigoplus_\ell \mathcal{Q}_\ell, 
\qquad
\mathcal Q_\ell = 
-\left(R_\ell + R_\ell^\dagger + 2 R_\ell^\dagger R_\ell\right) 
\otimes I_{2\ell+1}.
\]
We note that $\mathcal Q$ is HPSD iff $\mathcal R_\ell \ge 0  \; \forall \ell$, where we have defined:
\begin{equation}  \nonumber
\mathcal R_\ell \equiv 
-R_\ell - R_\ell^\dagger - 2 R_\ell^\dagger R_\ell = 
\begin{pmatrix} 
C_\ell & F_\ell \\ 
F_\ell^* & S_\ell 
\end{pmatrix}.
\end{equation}
The components of $\mathcal R_\ell$  can be calculated using the right-hand side of Eq.~\eqref{patti} and can be expressed in terms of $a_\ell$, $b_\ell$, $c_\ell$ as:
\be
C_\ell &=& 2 \left( \Re\{ a_\ell \} - |a_\ell|^2 - |c_\ell|^2 \right),\\
S_\ell &=& 2 \left(  \Re\{ b_\ell \} - |b_\ell|^2 - |c_\ell|^2 \right),\\ \label{FF_l}
F_\ell &=& 2 \left( \Im \{c_\ell\} - \im \left( b_\ell c_\ell^* - a_\ell^* c_\ell \right) \right).
\ee
Importantly, these $C_\ell$, $S_\ell$, and $F_\ell$ necessarily fulfill Sylvester's criterion, which can be compactly expressed as:
\begin{equation} \label{sylvesssr}
C_\ell \ge 0, \qquad
S_\ell \ge 0, \qquad
C_\ell S_\ell \ge |F_\ell|^2.
\end{equation}
The chiral Mie coefficients derived by Prof.~C.~F.~Bohren  in 1974~\cite{bohren1974light} satisfy the constraints given by Eq.~\eqref{sylvesssr}. To the best of our knowledge, these constraints have not been previously reported, despite the widespread use of $a_\ell$, $b_\ell$, and $c_\ell$ to study chiral light--matter interactions~\cite{ali2020enantioselective, ali2020probing, li2022enatioselective, wu2022selective, dutra2026circular, shi2020chirality, lai2024observation, feng2025revealing, miao2025long, shi2026self, diniz2025probing, olmos2024spheres, zheng2021selective}.

\begin{figure}[t!]
    \centering
\includegraphics[width=1\columnwidth]{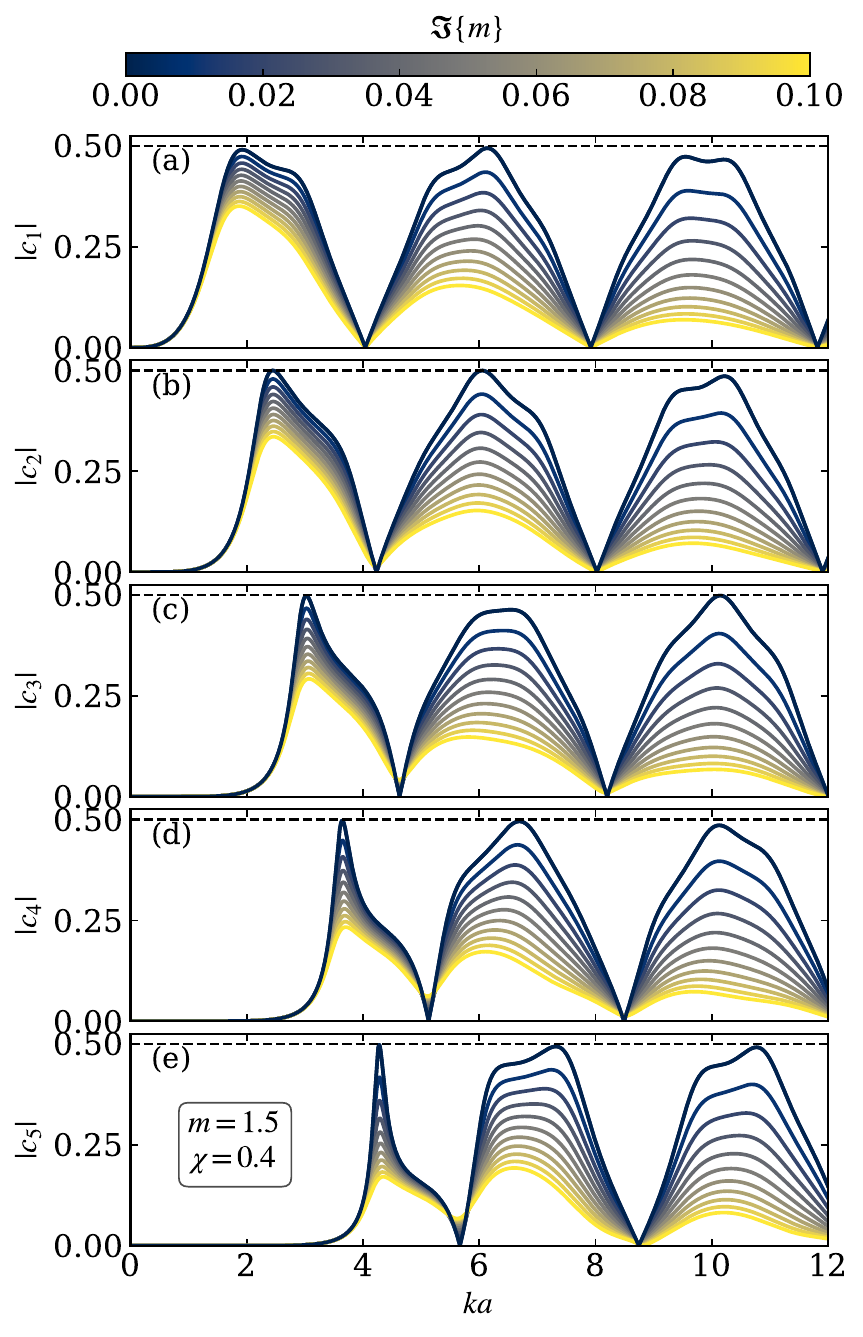}
   \caption{Chirality coefficient $|c_\ell|$ as a function of the size parameter $ka$ for the following fixed parameters: $\rm{m} = 1.5$ and $\chi = +0.4$. The horizontal dashed line indicates the theoretical upper bound of $0.5$.   The color gradient represents increasing material absorption $\Im\{m\} $.
    }
    \label{fig_1}
\end{figure}

\begin{figure}[t!]
    \centering
\includegraphics[width=1\columnwidth]{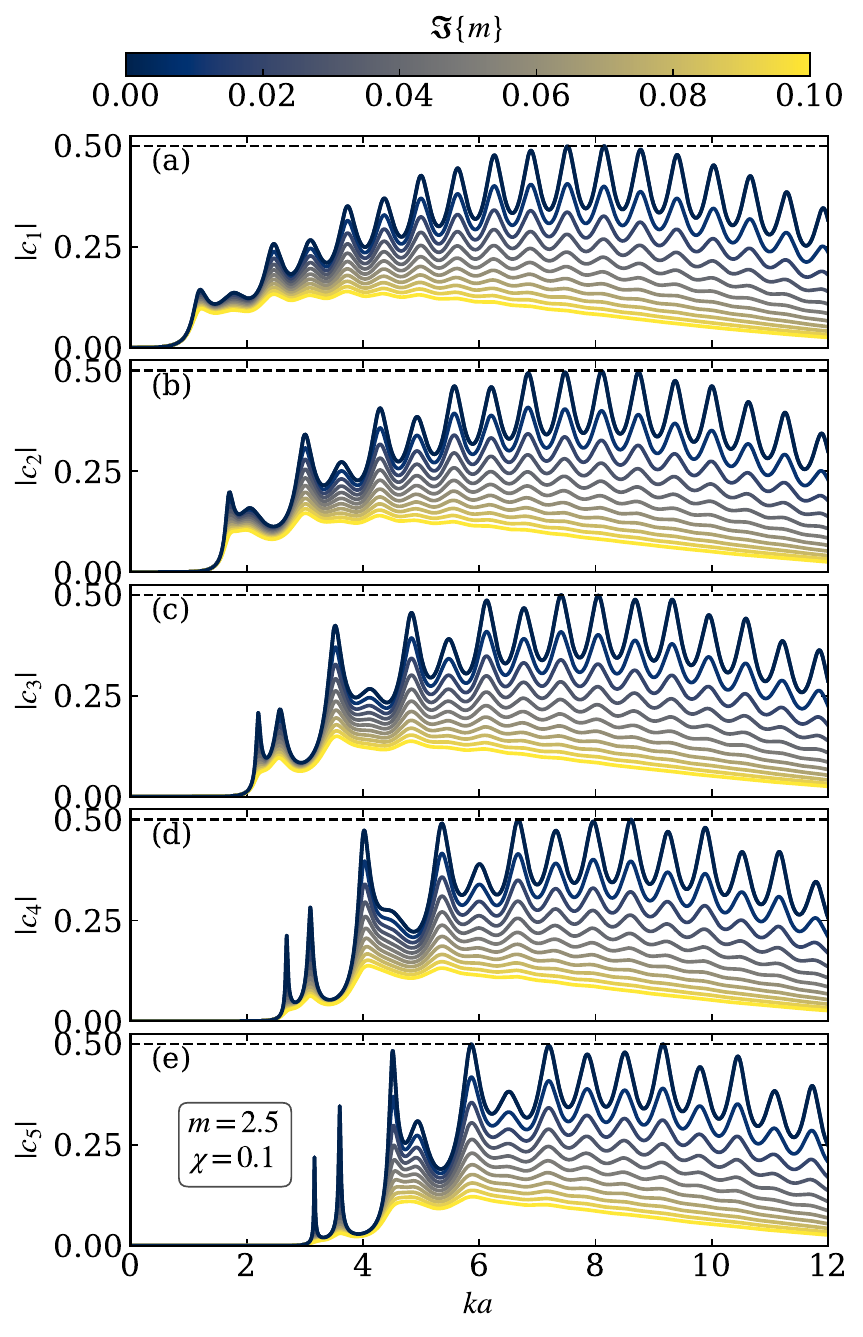}
    \caption{Chirality coefficient $|c_\ell|$ as a function of the size parameter $ka$ for the following fixed parameters: $\rm{m} = 3.5$ and $\chi = +0.1$. The horizontal dashed line indicates the theoretical upper bound of $0.5$.   The color gradient represents increasing material absorption $\Im\{m\}$.}
    \label{fig_2}
\end{figure}

However, the most important feature that can be derived from our novel inequalities is the ultimate upper bound of  $c_\ell$. 
Combining $C_\ell$ and $S_\ell$ with Eq.~\eqref{sylvesssr}  we arrive at:
\begin{equation} \nonumber
|c_\ell|^2 \le \Re \{ a_\ell \} - |a_\ell|^2, \qquad
|c_\ell|^2 \le \Re \{ b_\ell \} - |b_\ell|^2.
\end{equation}
We note that maximizing $|c_\ell|^2$ implies finding the maximum of: $
f(z) =\Re \{z - |z|^2 \} \equiv z' - z'^2 - z''^2.$
Here, $z$ stands for either $a_\ell$ or $b_\ell$;  
$z' \equiv \Re \{ z \}$, $z'' \equiv \Im \{z \}$.

Now, the function 
$f(z)=\Re\{z\}-|z|^2$ satisfies $f(z)\le 1/4$, where the equality is met only at $z=1/2$. 
Thus, we can write
\[
|c_\ell|^2 \le \Re\{a_\ell\}-|a_\ell|^2 \le \frac14,
\qquad
|c_\ell|^2 \le \Re\{b_\ell\}-|b_\ell|^2 \le \frac14.
\]
Saturation of the bound given by $|c^{\rm{max}}_\ell| = 1/2$ immediately gives 
$a_\ell= b_\ell=1/2$. 
This can be compactly expressed as
\begin{equation} \label{cultimate} 
|c^{\rm{\rm max}}_\ell| = \frac{1}{2} \Longrightarrow a_\ell = b_\ell = \frac{1}{2}.
\end{equation}
Equation~\eqref{cultimate} is the second important result of this work. The chiral scattering coefficient $c_\ell$ exhibits an identical upper bound for all multipolar orders $\ell$, independent of the optical size and specific material properties (including instrinsic chirality) of the spherical chiral object.
This bound constrains, for instance, the maximum attainable chirality transfer and circular dichroism enhancements in spherical chiral resonators under general illumination conditions~\cite{dutra2026circular}.

Importantly, when the upper bound of $c_\ell$  is reached, the object is necessarily non-absorbing in that channel. This holds because Eq.~\eqref{cultimate} implies $C_\ell = S_\ell = F_\ell = 0$. However, in this case, the converse statement does not hold: $C_\ell = S_\ell = F_\ell = 0$ does not generally yield Eq.~\eqref{cultimate}.

Next, we support our analytical findings by computing the chiral coefficient $c_\ell$ for a specific spherical chiral object experimentally synthesized in the seminal work of Ref.~\cite{shi2020chirality}.

As illustrated in Fig.~\ref{fig_1}, $|c_\ell|$ reaches its ultimate upper bound of $0.5$ for every multipolar order ($\ell = 1, 2, 3, 4, 5$) exclusively when the spherical chiral object is lossless. Any absorption, modeled as $\Im\{m\} > 0$ in Fig.~\ref{fig_1},  prevents the spherical chiral object from attaining the upper bound of $|c_\ell| $, 
regardless of its size parameter $ka$, real refractive index contrast $m$, or intrinsic chirality $\chi$. 
For completeness, Fig.~\ref{fig_2} shows the same behavior for alternative fixed values of $m$ and $\chi$. This supports that the upper bound of $|c_\ell|=0.5$ is a universal feature attainable only in the absence of optical losses.


\emph{Conclusions $\&$ Outlook.—--} We have found a novel upper bound in light--matter interactions: the cross-polarizabilities $\alpha_{\rm em}$ and $\alpha_{\rm me}$ of any bi-isotropic NP cannot exceed one half of the maximum value of the polarizability of a resonant isotropic point electric dipole.
Importantly, we have demonstrated that chiral or Tellegen NPs can reach the upper bound only if they are lossless.

We have further generalized our findings to spherical chiral objects of arbitrary optical size. Specifically, we have shown that the chiral Mie coefficient $c_\ell$ possesses an ultimate upper bound of $0.5$, which—just as in bi-isotropic NPs—can only be attained in the absence of losses.
Importantly, because the upper bound stems from energy conservation, it holds under general illumination conditions and for any specific choice of material parameters.


Ultimately, this work introduces a universal metric for the magnetoelectric coupling of bi-isotropic objects, quantifying how closely these can approach their ultimate upper bound.


\clearpage

\section*{References}
\bibliography{Bib_tesis}

\clearpage

\onecolumngrid

\end{document}